\documentclass
[prd,preprint,showpacs,groupedaddress]{revtex4-1}
\usepackage{amsmath}
\usepackage{amsfonts}
\usepackage{amssymb}
\usepackage{geometry}
\usepackage{natbib}
\usepackage[english]{babel}
\usepackage{graphicx}
\usepackage{epstopdf}
\begin{document}

  \title{Hawking-Page phase transition, Page curve and islands in black holes}

  \author{Dao-Quan Sun}
\email{lygsdq@outlook.com}
\affiliation{Lianyungang Normal University, Lianyungang 222006, P. R. China\\%
School of Physics, Sun Yat-sen University, Guangzhou 510275, P. R. China}

  \date{\today}

  \begin{abstract}

We study the Page curve and the effect of the Hawking-Page phase transition in the Vaidya model of evaporating BTZ black holes, both with and without islands. The phase transition can occur before, at, or after the Page time. When it occurs before the Page time, the information release does not require an island yet still reproduces the Page curve.
From the first law of entanglement, the entanglement temperature changes from negative to positive, diverging at the Page time. Thus, the information deposition phase corresponds to a negative temperature, and the release phase to a positive temperature.
If the Hawking-Page phase transition takes place, all black hole information is released at once via a quantum jump. We speculate that first-order phase transitions universally affect black hole information, providing an alternative escape mechanism independent of Hawking radiation.

  \end{abstract}


  \keywords { Hawking-Page transition, Page Curve, entanglement entropy.}

  \maketitle

\section{INTRODUCTION}

In 1975, Hawking~\cite{Hawking1975} showed that a black hole radiates like a black body with a temperature. This discovery resolved the temperature paradox in black hole thermodynamics and revealed an intrinsic connection among gravity, thermodynamics, and quantum mechanics. However, it also gave rise to the black hole information paradox. The discovery of the AdS/CFT correspondence by Maldacena~\cite{Maldacena:1997re} provided a framework to address whether information can escape a black hole. Within this correspondence, Ryu and Takayanagi~\cite{PhysRevLett.96.181602} proposed a holographic description of the entanglement entropy, which was later generalized and applied to the black hole information problem. The Page curve~\cite{Page:1993wv,Page:2013dx} predicts the time evolution of the entanglement entropy between the outgoing Hawking radiation and the remaining black hole, assuming that black hole evaporation is a unitary process. Reproducing the Page curve is a key step toward solving the information paradox. Recently, the Page curve for the entanglement entropy of Hawking radiation was successfully reproduced by developing the island formula~\cite{Almheiri:2019hni,Penington:2019npb,Almheiri:2019psf,Almheiri:2020cfm}. Including the island contributions, the fine-grained entropy of Hawking radiation is~\cite{Almheiri:2019hni}
\begin{equation}
\label{eq:island}
S(R)=\min\Bigl\{\operatorname{ext}\Bigl[\frac{\operatorname{Area}(\partial I)}{4G_{N}}+S_{\text{matter}}(R\cup I)\Bigr]\Bigr\},
\end{equation}
where $R$ and $I$ denote the radiation region and the island region, respectively, and $\partial I$ is the boundary of the island.
The island formula can be derived from the replica trick in gravitational backgrounds, where replica wormholes provide new saddles of the gravitational path integral, ultimately leading to the island formula~\cite{Almheiri:2019qdq,Penington:2019kki}.

For a sufficiently small subsystem, Ref.~\cite{PhysRevLett.110.091602} established a relation between the entanglement entropy of an excited state and its energy, which is analogous to the first law of thermodynamics. References~\cite{Blanco:2013joa,Wong:2013gua} further showed that, under a change in the quantum state, the entanglement entropy satisfies a thermodynamic-like relation
\begin{equation}
\label{eq:f-law}
\delta S_{EE}= \delta \langle H_{A} \rangle,
\end{equation}
where $S_{EE}$ is the entanglement entropy and $H_{A}$ is the modular Hamiltonian. More recently, some authors~\cite{PhysRevD.95.106007,PhysRevD.100.106008} generalized the concept of entanglement temperature using the first law of black hole thermodynamics. The entropy of a thermodynamic system is related to its internal energy via the first law of thermodynamics; similarly, the holographic entanglement entropy is also related to the internal energy of the system via this thermodynamic-like relation. This suggests that the energy of a black hole connects quantum information to its thermodynamic information. When the Hawking-Page phase transition occurs, the energy of the black hole undergoes a sudden change. This naturally raises the question of whether the Hawking-Page phase transition affects black hole information.

The Hawking-Page phase transition was first introduced by Hawking and Page~\cite{Hawking1983}, who demonstrated the existence of a phase transition between a thermal radiation phase and a stable black hole phase in an AdS spacetime. Since then, black hole phase transitions and critical phenomena have been extended to a variety of more complicated backgrounds~\cite{HENNEAUX1984415,1126-6708-1999-04-024,Witten:1998zw}. One key idea is that the cosmological constant $\Lambda$~\cite{HENNEAUX1984415,TEITELBOIM1985293} in an asymptotically AdS black hole spacetime can be interpreted as a pressure~\cite{0264-9381-26-19-195011} $P=-\frac{\Lambda}{8\pi G_{N}}$ in a thermodynamic sense.

Much of the early work on the information paradox focused on eternal black holes (see, e.g.,~\cite{Arefeva:2021kfx,Hashimoto:2020cas,Ling:2020laa,Chou:2021boq,Wang:2021woy,Matsuo:2020ypv,Saha:2021ohr,Yu:2021rfg}). More recently, attention has turned to understanding Hawking radiation in time-dependent backgrounds~\cite{Chou:2023adi,Guo:2023gfa}.

In this paper, we provide a simple toy model that describes the entire process from the gravitational collapse of a null shell forming a black hole of mass $M$ to its complete evaporation, returning to a pure AdS spacetime. We study the Page curve and the influence of the Hawking-Page phase transition on the Page curve for the spherically symmetric Vaidya model of evaporating BTZ black holes, both with and without island configurations. We also investigate the first law of entanglement in both the boundary and the bulk, and analyze the change of the entanglement temperature. Furthermore, we find that when the Hawking-Page phase transition occurs before or at the Page time, islands are not required to reproduce the Page curve. The entanglement temperature flips sign at the Page time, changing from negative (information deposition phase) to positive (information release phase). At the phase transition, all information stored in the black hole is released at once. We also study the effect of first-order phase transitions on black hole information, and speculate that this effect is universal.

The paper is organized as follows. In Sec.~\uppercase\expandafter{\romannumeral2}, we employ the Vaidya metric to construct spacetimes that contain evaporating BTZ black holes. In Sec.~\uppercase\expandafter{\romannumeral3}, we compute the entanglement entropy for this model, considering both scenarios with and without island configurations. In Sec.~\uppercase\expandafter{\romannumeral4}, we study the effect of the Hawking-Page phase transition on the Page curve. In Sec.~\uppercase\expandafter{\romannumeral5}, we study the first law of entanglement in both the boundary and the bulk. In Sec.~\uppercase\expandafter{\romannumeral6}, we present our conclusions and discussion. Throughout this paper, we set $\hbar=k_{B}=1$.

\section{The Vaidya model of an evaporating black hole}

In this section, we consider the Vaidya model of an evaporating BTZ black hole. We first review the BTZ black hole, whose metric can be written as
\begin{equation}
  \label{eq:background}
  ds^2 = -f(r)\,dt^2 + \frac{dr^2}{f(r)} + r^2 d\phi^2,
\end{equation}
with
\begin{equation}
  \label{eq:fr}
  f(r) = -8G_N M + \frac{r^2}{l^2} = \frac{r^2 - r_+^2}{l^2},
\end{equation}
and the cosmological constant $\Lambda = -1/l^2$. The parameter $r_+ = \sqrt{8G_N M}\,l$ is the horizon radius of the black hole. The relevant thermodynamic quantities are given by
\begin{align}
\label{eq:thermodynamic-q}
  T_{\mathrm{BH}} &= \frac{f'(r_+)}{4\pi} = \frac{r_+}{2\pi l^2}, \nonumber\\
  M &= \frac{r_+^2}{8G_N l^2}, \nonumber\\
  S_{\mathrm{BH}} &= \frac{\pi r_+}{2G_N}, \nonumber\\
  P &= \frac{1}{8\pi G_N l^2}, \nonumber\\
  V &= \pi r_+^2.
\end{align}
Here $T_{\mathrm{BH}}$ is the Hawking temperature at the horizon $r_+$, $S_{\mathrm{BH}}$ the thermodynamic entropy, $P$ the thermodynamic pressure, and $V$ the thermodynamic volume.

Much of the early work on the information paradox was devoted to the analysis of eternal AdS black holes. Here we provide a simple toy model describing the entire process from the gravitational collapse of a null shell that forms a black hole of mass $M_0$ to its complete evaporation, ultimately returning to a pure AdS spacetime.

\begin{figure}[htbp]
\centering
\includegraphics[width=12.2cm]{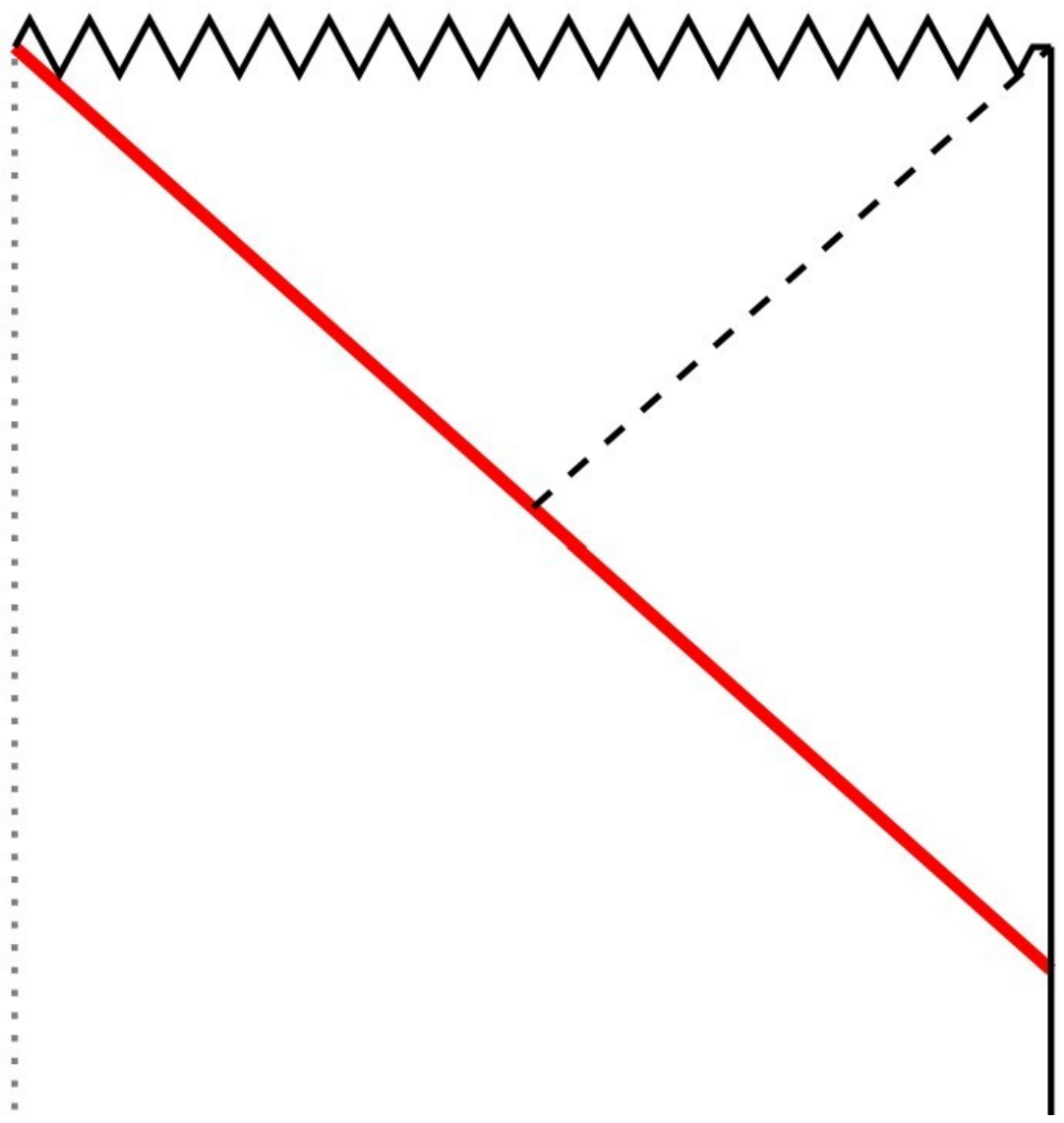}
\caption{
Penrose diagram for the AdS-Vaidya spacetime. The thick solid red line represents the infalling shell of matter in the zero-thickness limit. The left vertical gray dotted line represents the origin of spherical coordinates before the shell collapses, and the thick black wavy curve represents the singularity. On the right, the AdS boundary is represented by a thick solid black line; the black dashed line is the horizon of the black hole. The region above the thick solid red line corresponds to Schwarzschild-AdS spacetime, while the region below is pure AdS spacetime.}
\label{fig:1-1}
\end{figure}

The paper~\cite{Hubeny:2013dea} investigates the spherically symmetric Vaidya-AdS spacetime, which characterizes black hole formation due to the gravitational collapse of a null shell in the zero-thickness limit, thereby providing a simple toy model of AdS black hole formation. For BTZ black hole formation via the gravitational collapse of a null shell, we adopt the toy model introduced in Ref.~\cite{Hubeny:2013dea}. For an illustration of the black hole formation through the collapse of a null shell, see Fig.~\ref{fig:1-1}.

To investigate the evaporation process of a black hole, we consider a bulk geometry given by the three-dimensional outgoing Vaidya-AdS spacetime, with the metric
\begin{equation}
  \label{eq:vaidya-ads1}
  ds^2 = -f(r,u)\,du^2 - 2\,du\,dr + r^2 d\phi^2,
\end{equation}
where
\begin{equation}
  \label{eq:fr-vaidya}
  f(r,u) = \frac{r^2}{l^2} + 1 - \eta(u)\left(\frac{r_+^2}{l^2} + 1\right),
\end{equation}
and $\eta(u)$ is a monotonic function that decreases from $1$ in the past to $0$ in the future, characterizing an evaporating black hole model. Specifically, the black hole mass $M(u)$ is nonzero for $\eta(u)=1$ and vanishes for $\eta(u)=0$. Here $M(u)$ is an arbitrary decreasing function of $u$ satisfying $\lim_{u\to 0} M(u) = M_0 > 0$ and $\lim_{u\to u_0} M(u) = 0$.

\begin{figure}[htbp]
\centering
\includegraphics[width=12.2cm]{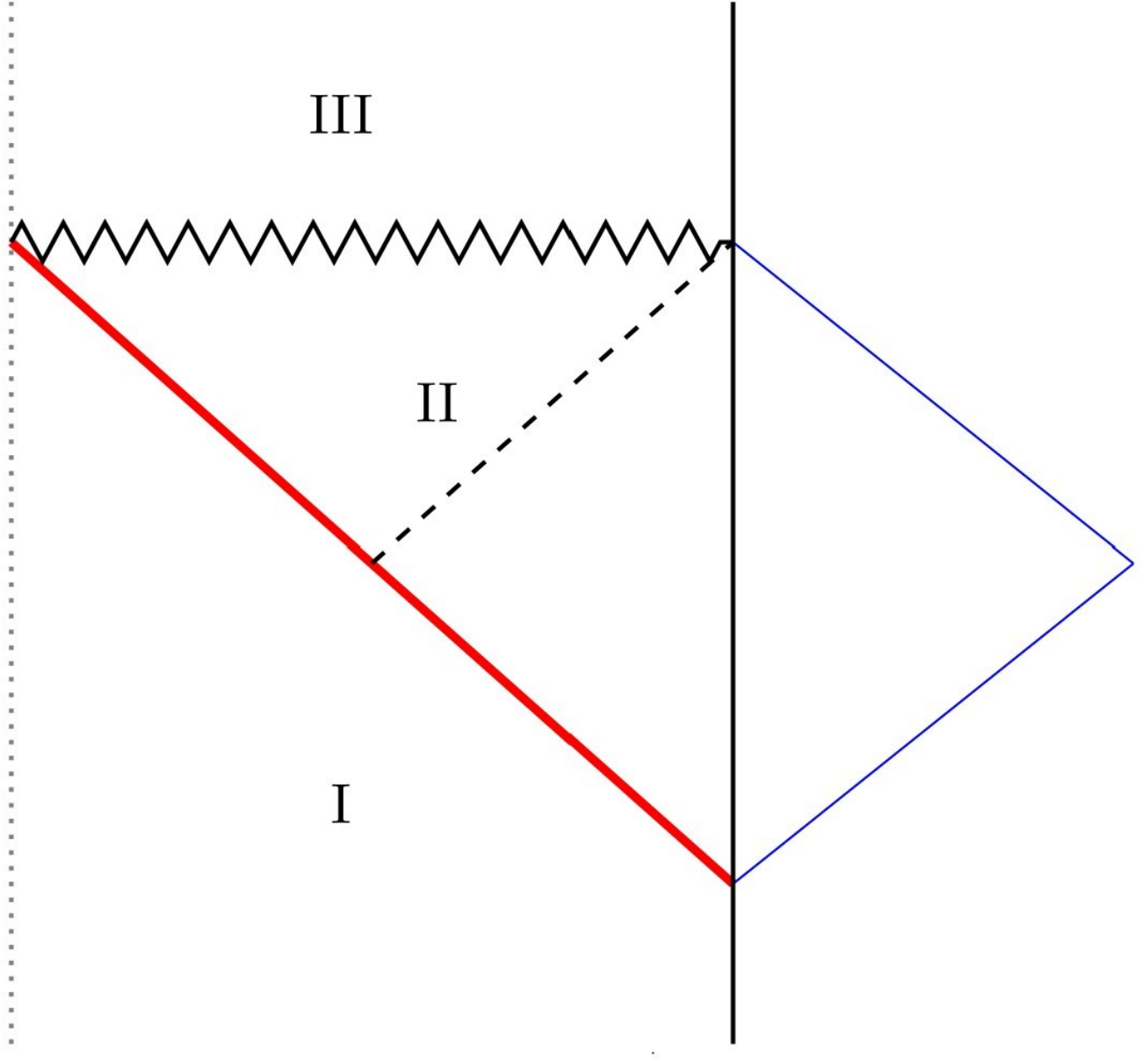}
\caption{
Penrose diagram for the AdS-Vaidya spacetime. The thick solid red line represents the infalling shell of matter in the zero-thickness limit. The left vertical gray dotted line represents the origin of spherical coordinates before the shell collapses, and the thick black wavy curve represents the singularity. On the right, the AdS boundary is represented by a thick solid black line, and an auxiliary flat thermal bath is attached to this boundary. The black dashed line is the horizon of the black hole. The region above the thick solid red line corresponds to Schwarzschild-AdS spacetime, while the region below is pure AdS spacetime. The uppermost region III is also pure AdS spacetime.}
\label{fig:1-2}
\end{figure}

To model the evaporation of a BTZ black hole, we adopt the standard technique of attaching auxiliary flat thermal baths~\cite{Almheiri:2013hfa,VanRaamsdonk:2013sza} to the gravitational region by matching them at the asymptotic boundary $r \to \infty$. To enable Hawking radiation to escape into the thermal baths, we impose a transparent boundary condition on the gravitational region. For an illustration of the model depicting the BTZ black hole from formation to evaporation, see Fig.~\ref{fig:1-2}. Regions I and III are pure AdS spacetimes, while region II is described by the Vaidya metric with a decreasing mass function.

The present model describes an evaporating black hole spacetime. For the period before the collapse, the geometry is initially pure AdS. At $u=0$, a null shell collapses in the zero-thickness limit, forming a BTZ black hole of mass $M_0$. Subsequently, the black hole begins to evaporate, and its evolution is governed by the outgoing AdS-Vaidya metric introduced in Eq.~\eqref{eq:vaidya-ads1}. At this stage, the boundary condition on the gravitational region becomes transparent, allowing Hawking radiation to escape into the thermal baths.

To keep the process tractable, we assume that the evaporation proceeds quasi-statically, i.e., at an extremely slow rate. This guarantees that the gravitational region stays in thermal equilibrium with the baths throughout the evaporation. The rate of mass loss is determined by the specific form of the decreasing function $M(u)$. As the evolution proceeds, the black hole mass gradually declines and eventually vanishes at $u=u_0$. In the final stage, the spacetime returns to a pure AdS configuration.

\section{Entanglement Entropy for the Vaidya Model of an Evaporating BTZ Black Hole}

In this section, we compute the entanglement entropy for three-dimensional evaporating BTZ black holes, examining both scenarios with and without island configurations. These results are then applied to analyze the Page curve within the context of the black hole information paradox, and to extract the corresponding Page time and scrambling time for our model.

To calculate the Page curve, we employ the Kruskal coordinates for the maximally extended BTZ spacetime, which cover both the interior and exterior regions. These coordinates are given by
\begin{equation}
\begin{aligned}
\label{eq:kr-c}
\text{Right Wedge}: U = -e^{-\kappa(t - r_{\ast})}, \quad V = e^{\kappa(t + r_{\ast})},\\
\text{Left Wedge}: U = e^{-\kappa(t - r_{\ast})}, \quad V = -e^{\kappa(t + r_{\ast})},
\end{aligned}
\end{equation}
where \(\kappa = \frac{2\pi}{\beta}\) is the surface gravity, and \(r_{\ast}\) denotes the tortoise coordinate defined by
\begin{equation}
\label{eq:r-star}
r_{\ast} = \int f^{-1}(r) dr = \frac{l^{2}}{2r_{+}} \log\left(\frac{|r - r_{+}|}{r + r_{+}}\right).
\end{equation}

In terms of the Kruskal coordinates, the metric becomes
\begin{equation}
\label{eq:metric-au}
ds^{2} = -\Omega^{-2} dU dV + r^{2} d\phi^{2},
\end{equation}
where the conformal factor is \(\Omega = \frac{r_{+}}{l(r + r_{+})}\).

For the entanglement entropy calculation, we consider a two-sided black hole system coupled to two auxiliary non-gravitational flat thermal baths. The evaporation is assumed to proceed quasi-statically, i.e., at an extremely slow rate, so that the gravitational region remains in thermal equilibrium with the baths throughout the process. A transparent boundary condition is imposed on the gravitational region, allowing Hawking radiation to escape into the thermal baths. For the three-dimensional BTZ black hole, the mass loss rate obeys the Stefan-Boltzmann law~\cite{Cardoso:2005cd,Xu:2019krv,Ong:2014maa}, yielding
\begin{equation}
\label{eq:S-B-mT}
\frac{dM}{du} = -\sigma T_{BH}^{3},
\end{equation}
where \(\sigma\) is a positive constant determined by the radiation species and the AdS scale. Combining Eq.~\eqref{eq:S-B-mT} with the thermodynamic relation Eq.~\eqref{eq:thermodynamic-q}, we obtain
\begin{equation}
\label{eq:S-B-m-u}
\frac{dM}{du} = -\sigma \left(\frac{\sqrt{2G_{N}}}{\pi l}\right)^{3} M^{\frac{3}{2}}.
\end{equation}
Consequently, the decreasing mass function \(M(u)\) takes the explicit form
\begin{equation}
\label{eq:S-B-m-m0}
M(u) = \left(M_{0}^{-1/2} + \frac{\sigma (2G_{N})^{\frac{3}{2}}}{2\pi^{3} l^{3}} u\right)^{-2}.
\end{equation}
This expression gives \(M(0)=M_{0}\) and \(M(u)\to 0\) only as \(u\to\infty\). In practice, after a sufficiently long time \(u=u_{0}\) (which can be regarded as the evaporation timescale), the mass becomes negligibly small, i.e., \(M(u_{0})\approx 0\). Therefore, we may approximate the evaporation process as ending at a finite time \(u_{0}\) by effectively taking the limit \(u_{0}\to\infty\). This guarantees that the spacetime returns to a nearly pure AdS configuration after a long enough evolution.

\begin{figure}[htbp]
\centering
\includegraphics[width=12.2cm]{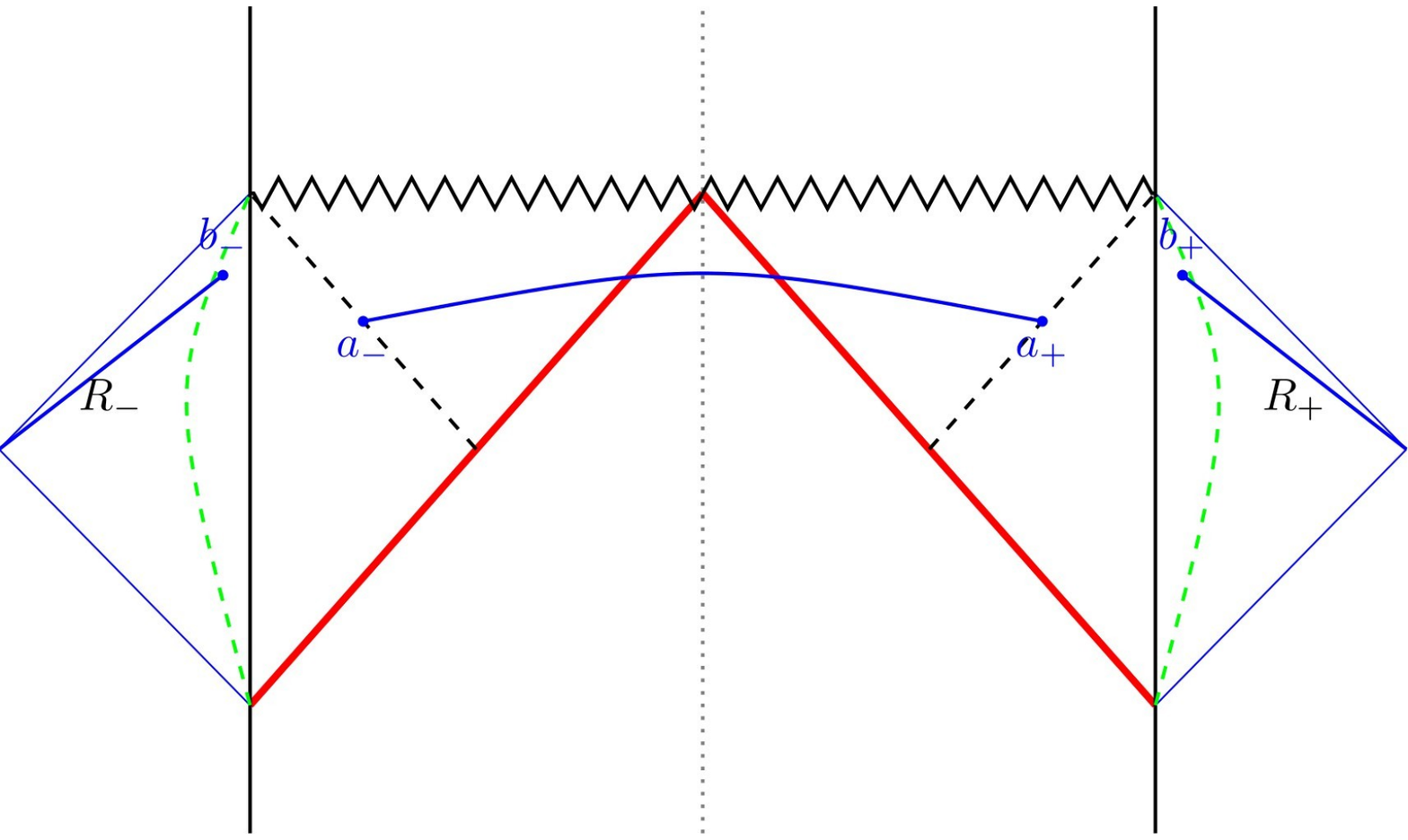}
\caption{
Penrose diagram for the evaporation model of the two-sided black hole system. The thick solid red line represents the infalling shell of matter in the zero-thickness limit. The middle vertical gray dotted line represents the origin of spherical coordinates before the shell starts, and the thick black wavy curve represents the singularity. The AdS boundary is represented by a thick solid black line, and an auxiliary flat thermal bath is attached to this boundary. The black dashed line is the horizon of the black hole. The region above the thick solid red line corresponds to Schwarzschild-AdS spacetime, while the region below it is pure AdS spacetime. The uppermost region is also pure AdS spacetime. We denote \(a_{\pm}=(\pm u_{a},a)\) and \(b_{\pm}=(\pm u_{b},b)\).}
\label{fig:1-3}
\end{figure}

From Eqs.~\eqref{eq:S-B-m-m0} and \eqref{eq:thermodynamic-q}, we obtain the explicit form of the temperature function \(T_{BH}(u)\):
\begin{equation}
\label{eq:T-u}
T_{BH}(u) = \frac{\sqrt{2G_{N}}}{\pi l} \left( \frac{\sqrt{2G_{N}}}{\pi l} \frac{1}{T_{0}} + \frac{\sigma (2G_{N})^{\frac{3}{2}}}{2\pi^{3} l^{3}} u \right)^{-1},
\end{equation}
which satisfies \(T_{BH}(0)=T_{0}=\frac{\sqrt{2G_{N}M_{0}}}{\pi l}\) and tends to zero as \(u \to u_{0}\) (the final stage of evaporation). An illustration of the evaporation model for the two-sided black hole system is shown in Fig.~\ref{fig:1-3}.

In the subsequent analysis, we follow the approach of Refs.~\cite{Hashimoto:2020cas,Polchinski:2016hrw,Hubeny:2009rc} by considering only the s-wave of the conformal matter. This mode provides the dominant contribution and is equivalent to considering only massless modes. Under this reduction, the s-wave approximation is valid and the matter field theory simplifies to an effective two-dimensional CFT. Consequently, the entanglement entropy of Hawking radiation can be approximately expressed by the von Neumann entropy of quantum matter in the \(2d\) CFT.

We assume that the entire system is in a pure state and is filled with conformal matter of central charge \(c\). In the approximation described above, the matter sector is therefore described by an effective \(2d\) CFT. When the island does not appear, the entanglement entropy of quantum matter on \(R = R_{+} \cup R_{-}\) satisfies \(S_{\text{gen}}(R)=S(R)=S_{\text{matter}}(R)\), as can be seen from Eq.~\eqref{eq:island}. The generalized entanglement entropy of the quantum matter is then given by
\begin{equation}
\label{eq:s-r-b}
S_{\text{gen}}(R) = \frac{c}{3} \log\bigl[d(b_{+},b_{-})\bigr],
\end{equation}
where
\(d(b_{+},b_{-}) = \sqrt{\frac{[U(b_{-})-U(b_{+})][V(b_{+})-V(b_{-})]}{\Omega(b_{+})\Omega(b_{-})}}\)
denotes the distance between the cutoff boundaries \(b_{+}\) and \(b_{-}\). The precise form of this distance function is provided in Ref.~\cite{Saha:2021ohr}.

Equation~\eqref{eq:s-r-b} leads to
\begin{equation}
\label{eq:s-before-1}
S_{\text{gen}}(R) = \frac{c}{3} \log\left[ \left(\frac{1}{T_{BH}\pi l}\right) \sqrt{b^{2} - r_{+}^{2}} \, \cosh(2T_{BH}\pi u) \right],
\end{equation}
where \(T_{BH}\) is the temperature of the black hole and \(c\) is the central charge.

Now we set the order of \(b\) to be the same as that of \(r_{+}\). At late times \(u \gg \kappa\), we have \(\cosh(2T_{BH}\pi u) \approx \exp(2T_{BH}\pi u)\). Then Eq.~\eqref{eq:s-before-1} approximates to
\begin{equation}
\label{eq:s-before-3}
S_{\text{gen}}(R) \approx \frac{2c\pi^{3}l^{3}}{3\sigma G_{N}} \left(1 - \frac{1}{1 + \frac{\sigma T_{0}G_{N}}{\pi^{2}l^{2}} u}\right) = \left(\frac{2cG_{N}}{3\pi l^{4}} u\right) S_{BH}.
\end{equation}
From Eq.~\eqref{eq:s-before-3}, we see that when \(u > \frac{3\pi l^{4}}{cG_{N}}\), the entanglement entropy of the Hawking radiation becomes much greater than the thermodynamic entropy of the black hole at late times. To resolve this paradox, we need to take into account the contribution of the island. Using the formula for the entanglement entropy of two disconnected intervals~\cite{Calabrese:2009ez,Casini:2009sr},
\begin{equation}
\label{eq:s-after-0}
S_{\text{matter}}(R\cup I) = \frac{c}{3} \log\left[ \frac{d(a_{+},a_{-})\, d(b_{+},b_{-})\, d(a_{+},b_{+})\, d(a_{-},b_{-})}{d(a_{+},b_{-})\, d(a_{-},b_{+})} \right],
\end{equation}
where \(a_{\pm}=(\pm u_{a},a)\) and \(b_{\pm}=(\pm u_{b},b)\). Since the distances between the left and right wedges are very large, we have
\begin{equation}
\label{eq:d-d}
d(a_{+},a_{-}) \simeq d(b_{+},b_{-}) \simeq d(a_{\pm},b_{\mp}) \gg d(a_{\pm},b_{\pm}).
\end{equation}
Thus, after the Page time, the generalized entropy becomes
\begin{equation}
\label{eq:gen}
S_{\text{gen}}(R) \approx \frac{\pi a}{G_{N}} + \frac{c}{3} \log\bigl[d(a_{+},b_{+})\bigr].
\end{equation}

The extremization condition \(\frac{\partial S_{\text{gen}}}{\partial u_{a}} = 0\) yields \(u_{a}=u_{b}\). Then the extremization condition \(\frac{\partial S_{\text{gen}}}{\partial a} = 0\) gives
\begin{equation}
\label{eq:-a}
a \approx r_{+}.
\end{equation}
Substituting these extremized values into \(S_{\text{gen}}\) yields
\begin{equation}
\label{eq:-s-gen}
S_{\text{gen}}(R) \approx 2S_{BH} = 2\sqrt{\frac{2}{G_{N}}}\,\pi l^{3}\left(\frac{\sqrt{2G_{N}}}{\pi l}\frac{1}{T_{0}}+\frac{\sigma(2G_{N})^{\frac{3}{2}}}{2\pi^{3}l^{3}}u\right)^{-1}.
\end{equation}

\begin{figure}[htbp]
\centering
\includegraphics[width=10.2cm]{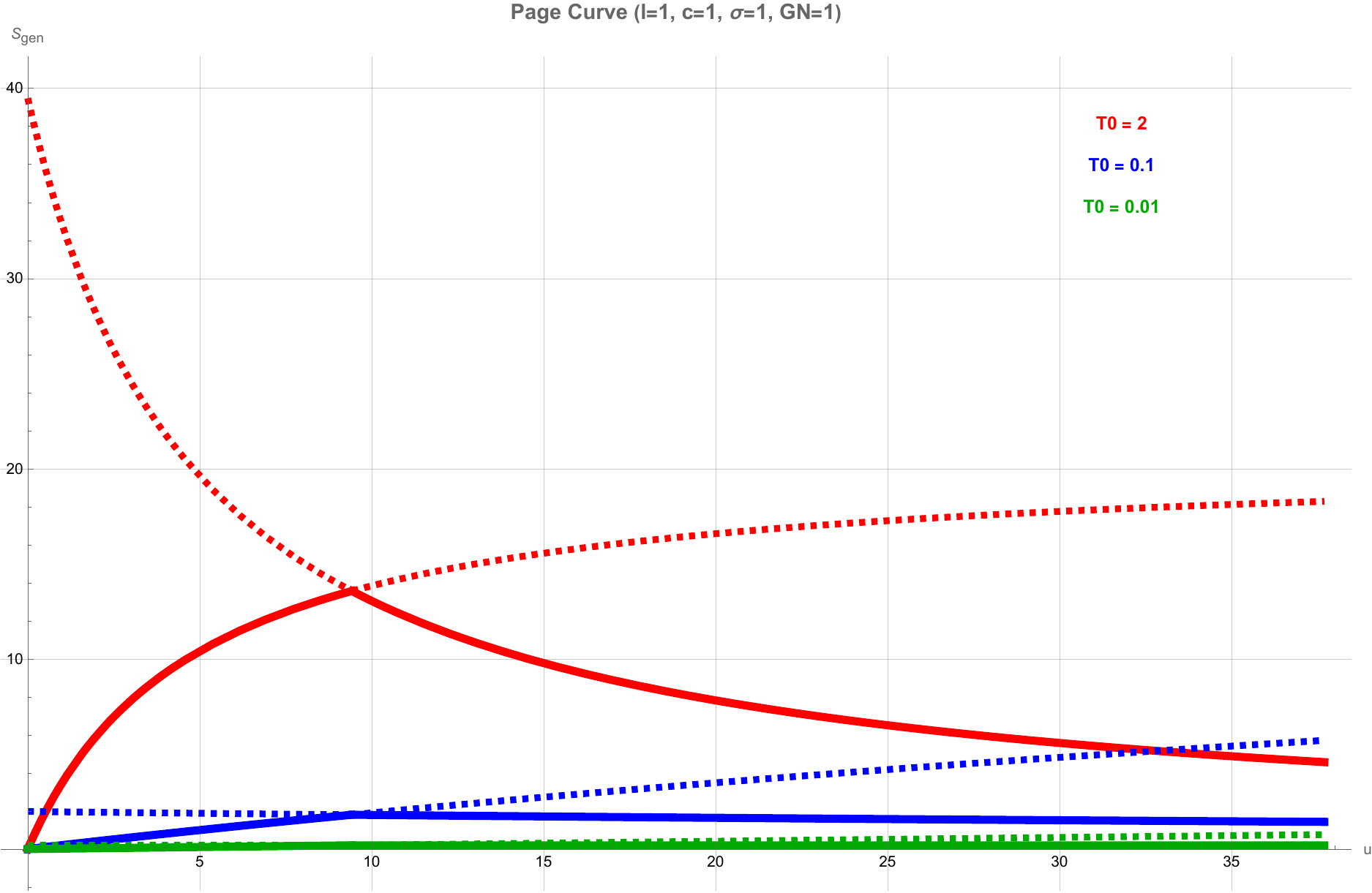}
\caption{The Page curve without a Hawking-Page phase transition.}
\label{fig:1}
\end{figure}

The entanglement entropy of Hawking radiation decreases at late times due to the contribution of the island. Thus, this gives the correct Page curve and therefore resolves the information loss paradox. Here we assume that no Hawking-Page phase transition occurs during the entire evaporation process of the black hole, and the Page curve is illustrated in Fig.~\ref{fig:1}.

From Eqs.~\eqref{eq:s-before-3} and \eqref{eq:-s-gen}, we obtain the Page time
\begin{equation}
\label{eq:t-page}
u_{\text{Page}} \approx \left(\frac{3}{c \pi T_{BH}}\right) S_{BH} = \frac{3\pi l^{2}}{c G_{N}}.
\end{equation}
From Eq.~\eqref{eq:t-page}, we see that \(u_{\text{Page}} < \frac{3\pi l^{4}}{c G_{N}}\); that is, the Page time occurs before the entanglement entropy of the Hawking radiation exceeds the thermodynamic entropy of the black hole.

The scrambling time~\cite{Hayden:2007cs} is defined as the shortest time required to recover information from the Hawking radiation. It also corresponds to the time for information to enter the surface of the island:
\begin{equation}
\label{eq:t-scr}
\begin{aligned}
u_{\text{scr}} & \equiv r_{\ast}(b) - r_{\ast}(a) \approx \frac{1}{2\pi T_{BH}} \log(S_{BH}) \\
&= \frac{l}{2\sqrt{2G_{N}}} \left( \frac{\sqrt{2G_{N}}}{\pi l}\frac{1}{T_{0}} + \frac{\sigma(2G_{N})^{\frac{3}{2}}}{2\pi^{3}l^{3}} u \right) \\
&\quad \times \log\left( \sqrt{\frac{2}{G_{N}}} \pi l^{3} \left( \frac{\sqrt{2G_{N}}}{\pi l}\frac{1}{T_{0}} + \frac{\sigma(2G_{N})^{\frac{3}{2}}}{2\pi^{3}l^{3}} u \right)^{-1} \right).
\end{aligned}
\end{equation}
From Eq.~\eqref{eq:t-scr}, we see that the scrambling time varies with time \(u\).

\section{The effect of the Hawking-Page phase transition on the Page Curve}

In the previous study of the BTZ black hole evaporation process, the possible occurrence of the Hawking-Page phase transition was not considered. In this section, we explore the situation where such a phase transition takes place during the black hole evaporation.

The paper~\cite{Dolan:2010ha} shows that a first-order Hawking-Page phase transition occurs for the BTZ black hole, causing it to transition to pure AdS\(_3\) spacetime when the Gibbs free energies of the two phases become equal. The relation between the thermodynamic temperature and the mass of the black hole is given by
\begin{equation}
T_{BH} = \frac{\sqrt{2G_N M}}{\pi l}.
\end{equation}
This shows that the thermodynamic temperature decreases as the black hole loses mass via Hawking radiation. Since the heat capacity \(C_p > 0\), there is no local instability. Consequently, the Hawking-Page phase transition can occur through the emission of Hawking radiation. When the black hole temperature reaches the critical value \(T_{HP} = \frac{1}{2\pi l}\) and the critical mass \(M_{HP} = \frac{1}{8G_N}\), the Hawking-Page phase transition takes place.

In our evaporating model, the black hole first loses mass via Hawking radiation according to Eq.~\eqref{eq:S-B-m-m0}. When the mass decreases to the critical value \(M_{HP} = \frac{1}{8G_N}\) at \(u = u_{HP}\), the system undergoes a first-order Hawking-Page phase transition. In this work, we adopt the instantaneous transition approximation for this phase transition. This approximation is justified when the characteristic timescale of the phase transition, \(\tau\), is much shorter than the evaporation timescale, i.e., \(\tau \ll u_{HP}\).

The characteristic timescale of the phase transition can be estimated as the light-crossing time of the AdS space, \(\tau \sim l\), because bubble nucleation and expansion are causally limited to propagate at most at the speed of light. In our evaporation model, the phase transition time \(u_{HP}\) is determined by the condition \(M(u_{HP}) = M_{HP}\), yielding
\begin{equation}
\label{eq:u-Hp}
u_{HP} = \frac{\pi^3 l^3}{\sigma G_N} \left( 2 - \frac{1}{\pi l T_0} \right).
\end{equation}
From Eq.~\eqref{eq:u-Hp}, we see that the first term dominates in the semiclassical regime. Moreover,
\[
\frac{\tau}{u_{HP}} \sim \frac{G_N}{l^2} \ll 1,
\]
where the inequality holds for \(l \gg \sqrt{G_N}\), which is precisely the regime where the semiclassical treatment is valid. Therefore, the phase transition occurs on a timescale negligible compared to the evaporation process, and the instantaneous transition approximation is well justified.

Thus, while a realistic first-order phase transition involves a finite duration, the instantaneous approximation captures the essential physics — the abrupt release of information trapped in the black hole — without introducing unnecessary complexity. The mass function is therefore written as a discontinuous jump at \(u = u_{HP}\):
\begin{equation}
\label{eq:u-cc+}
M(u_{HP}^{-}) = \frac{1}{8G_N}, \qquad M(u_{HP}^{+}) = -\frac{1}{8G_N}.
\end{equation}
After the transition, the spacetime settles into pure AdS\(_3\) with a constant mass \(M(u_{HP}^{+}) = -\frac{1}{8G_N}\) for all \(u > u_{HP}\). The complete mass evolution is summarized as
\begin{equation}
\label{eq:summa}
M(u) =
\begin{cases}
\left( M_0^{-1/2} + \dfrac{\sigma (2G_N)^{3/2}}{2\pi^3 l^3} u \right)^{-2}, & 0 \le u < u_{HP}, \\[6pt]
-\dfrac{1}{8G_N}, & u > u_{HP}.
\end{cases}
\end{equation}
At the transition point, the information previously stored in the black hole is abruptly released into the thermal baths. This provides a natural mechanism for resolving the information paradox within the AdS/CFT correspondence and aligns with the Page curve expectations~\cite{Page:1993wv,Page:2013dx}. A schematic illustration of this evaporation process — including the formation, Hawking radiation stage, instantaneous phase transition, and final pure AdS spacetime — is shown in Fig.~\ref{fig:1-3}.

\begin{figure}[htbp]
\centering
\includegraphics[width=10.2cm]{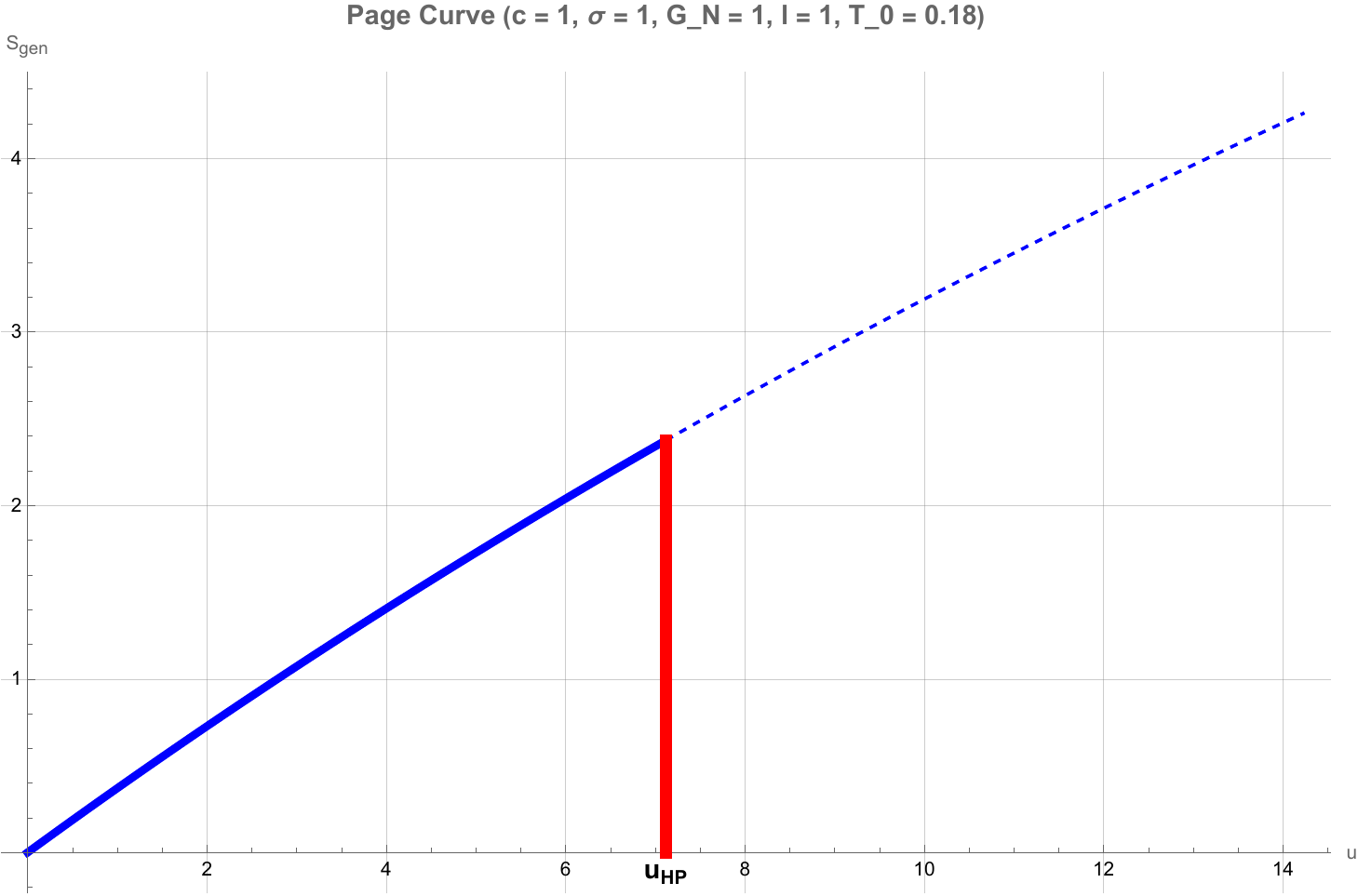}
\caption{The Page curve when the Hawking-Page phase transition occurs before the Page time.}
\label{fig:2}
\end{figure}

Now let us consider the effect of the Hawking-Page phase transition on the Page curve. If the Hawking-Page phase transition does not happen, the Page curve is the one illustrated in Fig.~\ref{fig:1}. If the phase transition occurs exactly at the Page time, i.e., \(u_{HP} = u_{\text{Page}}\), we can determine the initial mass \(M_0\) of the black hole from Eqs.~\eqref{eq:t-page} and \eqref{eq:u-Hp}. This initial mass can be regarded as a critical initial mass:
\begin{equation}
\label{eq:m-c}
M_c = \frac{1}{2G_N} \left( \frac{c\pi^2 l}{2c\pi^2 l - 3\sigma} \right)^2.
\end{equation}
The entanglement entropy of the Hawking radiation increases monotonically with time before the Page time. If the initial mass \(M_0\) satisfies \(\frac{1}{8G_N} < M_0 < M_c\), then the Hawking-Page phase transition occurs before the Page time. The corresponding Page curve is shown in Fig.~\ref{fig:2}, where the phase transition happens at \(u = u_{HP}\).

We observe that when the Hawking-Page phase transition takes place before or exactly at the Page time, the shape of the Page curve is as depicted in Fig.~\ref{fig:2}, with \(u_{\text{Page}} = u_{HP}\). At that moment, the thermal entropy of the black hole becomes
\begin{equation}
\label{eq:-s-hp}
S_{BH}^{(HP)} = \frac{\pi l}{2 G_N}.
\end{equation}
The Page curve drops rapidly to zero at the Hawking-Page transition point. This drop is much faster than the decrease of the Page curve due solely to black hole evaporation. Moreover, this process does not require the appearance of islands and yet reproduces the Page curve behavior.

\begin{figure}[htbp]
\centering
\includegraphics[width=10.2cm]{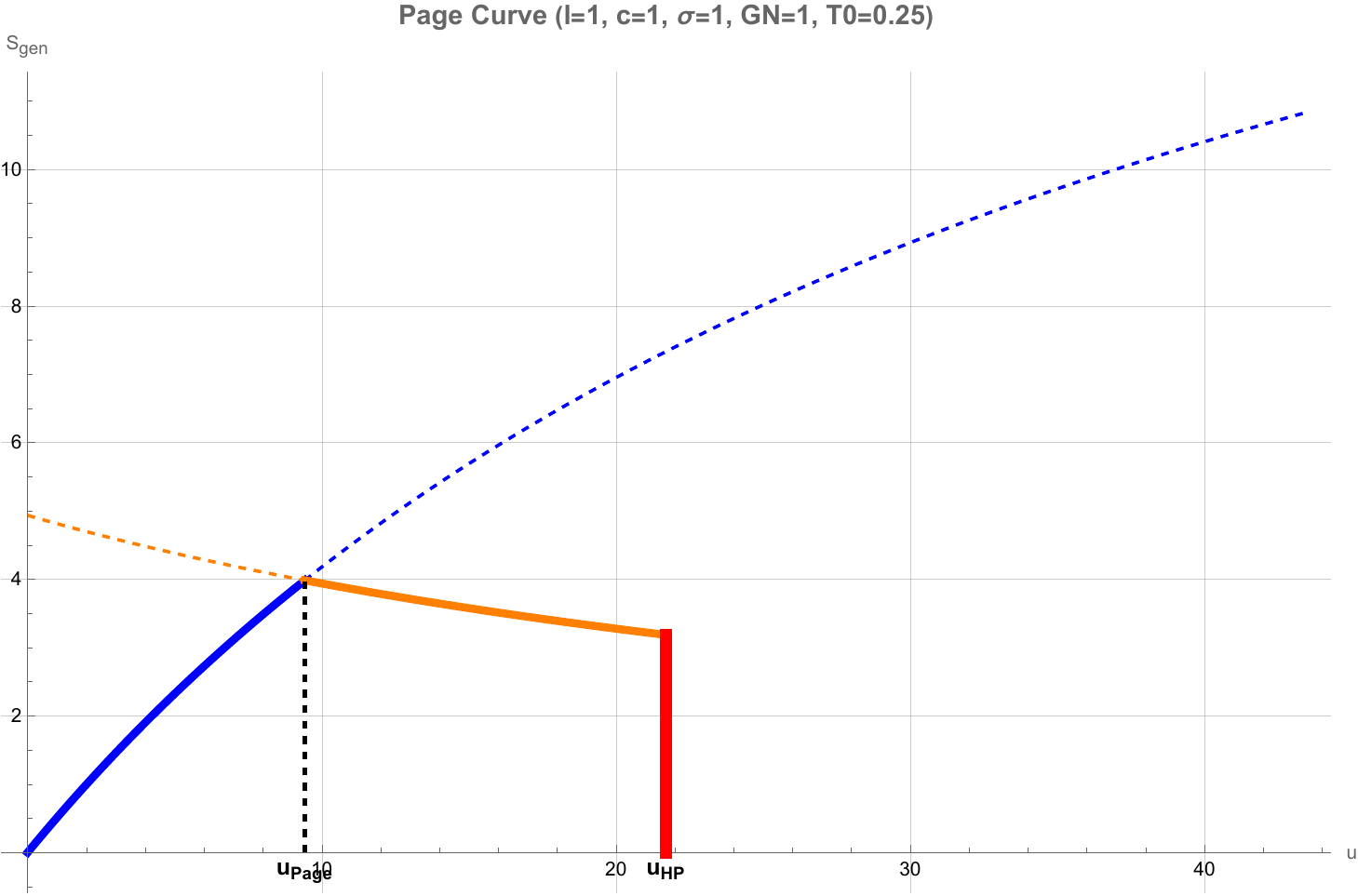}
\caption{The Page curve when the Hawking-Page phase transition occurs after the Page time.}
\label{fig:3}
\end{figure}

If the initial mass \(M_0\) satisfies \(M_0 > M_c\), then the Hawking-Page phase transition occurs after the Page time. In this case, the entanglement entropy of the Hawking radiation is bounded by the black hole entropy and becomes a decreasing function of time after the Page time. The quantum extremal surface is then located at
\begin{equation}
\label{eq:a-l}
a \approx r_{+} = l.
\end{equation}
The corresponding Page curve is illustrated in Fig.~\ref{fig:3}, where the Hawking-Page phase transition happens at \(u = u_{HP}\).

We observe that when the phase transition occurs after the Page time, the Page curve still drops rapidly to zero at the Hawking-Page transition point, much faster than the evaporation-induced decrease alone.

\section{The first law of entanglement and the Hawking-Page transition}

In this section, we study the first law of entanglement in the system. We first calculate the change of holographic entanglement entropy when the Hawking-Page phase transition occurs for a spherical region in this geometry, using the AdS/CFT correspondence.

The function \(f(r)\) in Eq.~\eqref{eq:fr} encodes the excitation properties of the boundary CFT. For a pure AdS spacetime, the three-dimensional function \(f(r)\) is
\begin{equation}
f(r)=1+\frac{r^2}{l^2},
\end{equation}
which corresponds to the ground state of the two-dimensional boundary CFT. On the boundary, we divide the total system into subsystems \(A\) and \(B\). The entanglement entropy \(S_A\) of subsystem \(A\) is defined as \(S_A = -\operatorname{Tr}\rho_A \log \rho_A\), where \(\rho_A = \operatorname{Tr}_B \rho_{\text{total}}\) is the reduced density matrix obtained by tracing out \(B\) from the total density matrix. In the gravity dual, the holographic entanglement entropy is given by~\cite{PhysRevLett.96.181602}
\begin{equation}
S_A = \frac{\operatorname{Area}(\gamma_A)}{4G_N},
\end{equation}
where \(\gamma_A\) is the minimal-area surface extending into the bulk.

We assume the system is on a fixed time slice and choose the entangling surface of subsystem \(A\) to be \(\theta = \theta_c\) with \(0 < \theta_c < \frac{\pi}{2}\). The corresponding bulk surface is described by \(r(\theta)\), with the minimum radius at \(\theta=0\) and \(r \to \infty\) at \(\theta = \theta_c\). Its area is then
\begin{equation}
\operatorname{Area} = 2\pi \int_0^{\theta_c} d\theta \sqrt{r^2(\theta) + \frac{r'^2(\theta)}{f(r(\theta))}}.
\end{equation}
For the pure AdS spacetime, the solution to this variational problem is
\begin{equation}
r^2(\theta) = \frac{l^2 \cos^2\theta_c}{\sin^2\theta_c - \sin^2\theta},
\end{equation}
and the holographic entanglement entropy \(S_A^{(0)}\) is~\cite{Calabrese_2004,Hubeny:2013gta}
\begin{equation}
S_A^{(0)} = \frac{c}{3} \log\left( \frac{2l}{\varepsilon} \sin\theta_c \right),
\end{equation}
where \(\varepsilon\) is a short-distance cutoff on the boundary, and \(c = \frac{3l}{2G_N}\) is the central charge of the two-dimensional boundary CFT.

For the BTZ black hole background, the holographic entanglement entropy reads~\cite{Hubeny:2012wa,Hubeny:2013gta}
\begin{equation}
S_A = \frac{c}{3} \log\left( \frac{\sinh(2\pi T_{BH} l \theta_c)}{\varepsilon \pi T_{BH}} \right).
\end{equation}

Now we consider that the Hawking-Page phase transition occurs for the BTZ black hole. Then the holographic entanglement entropy becomes
\begin{equation}
S_A^{(c)} = \frac{c}{3} \log\left( \frac{2l}{\varepsilon} \sinh \theta_c \right).
\end{equation}
The reduction of holographic entanglement entropy from the excited state to the ground state of the CFT in subsystem \(A\) is
\begin{equation}
\delta S_A^{(c)} = \frac{c}{3} \log\left( \frac{\sinh \theta_c}{\sin\theta_c} \right).
\end{equation}

When the Hawking-Page phase transition takes place, the reduced energy in subsystem \(A\) can be written as
\begin{equation}
\delta E_A^{c} = \frac{\theta_c}{8\pi G_N}.
\end{equation}
The first-law-like relation for a sufficiently small subsystem is
\begin{equation}
\delta E_A^{c} = T_{\text{ent}} \, \delta S_A^{(c)},
\end{equation}
where the entanglement temperature is
\begin{equation}
\label{eq:Tent}
T_{\text{ent}} = k \cdot \theta_c^{-1},
\end{equation}
with \(k\) an order one constant.  This result agrees with Ref.~\cite{PhysRevLett.110.091602}. From the above we see that when the Hawking-Page phase transition occurs, the boundary CFT undergoes a transition from an excited state to the ground state, so there is an energy change \(\Delta E_A^{c}\) that is related to the amount of entanglement entropy.

The BTZ black hole is dual to a thermofield double state in a two-dimensional CFT, while the matter field theory in the bulk can be approximated by an effective 2d CFT. In what follows, we investigate the first law of entanglement for the BTZ black hole in the bulk.

When the black hole initially begins to evaporate, its mass decreases. Let the energy change due to Hawking radiation be \(\delta E\) and the change in entanglement entropy be \(\delta S_{\text{gen}}(R)\). According to the first law of entanglement, \(\delta E = T_{\text{ent}} \delta S_{\text{ent}}\), we obtain the entanglement temperature
\begin{equation}
\label{eq:Te-pag-1}
T_{\text{ent}}(u) = -\frac{3\sigma G_N}{c\pi^3 l^3 \sqrt{M_0}}
\frac{\left(1 + \frac{\sigma\sqrt{2M_0}\, G_N^{3/2}}{\pi^3 l^3}u\right)^2}
{\left(M_0^{-1/2} + \frac{\sigma(2G_N)^{3/2}}{2\pi^3 l^3}u\right)^3} < 0.
\end{equation}
Before the Page time, we have \(\delta E < 0\) and \(\delta S_{\text{gen}}(R) > 0\), which gives a negative entanglement temperature \(T_{\text{ent}} < 0\); moreover, \(T_{\text{ent}}\) decreases with time \(u\). The Page curve also implies that the entanglement temperature must be negative before the Page time.

If the Hawking-Page phase transition occurs before the Page time (\(M_0 < M_c\)), then at the phase transition time \(u_{HP}\), under the instantaneous transition approximation, \(\delta S_{\text{gen}}(R)\) tends to negative infinity, and consequently the entanglement temperature tends to zero: \(T_{\text{ent}} \to 0\) at \(u = u_{HP}\).

If the Hawking-Page phase transition occurs after the Page time (\(M_0 > M_c\)), at the Page time \(u_{\text{Page}}\) the ratio \(\delta E / \delta S_{\text{gen}}(R)\) diverges, so \(T_{\text{ent}}\) diverges at \(u_{\text{Page}}\).

After the Page time, we have \(\delta E < 0\) and \(\delta S_{\text{gen}}(R) < 0\), which yields a positive entanglement temperature
\begin{equation}
\label{eq:Te-aft2}
T_{\text{ent}}(u) \approx T_{BH}(u) = \frac{\sqrt{2G_N}}{\pi l}
\left( \frac{1}{\sqrt{M_0}} + \frac{\sigma (2G_N)^{3/2}}{2\pi^3 l^3} u \right)^{-1} > 0.
\end{equation}
At this stage, the entanglement temperature is approximately equal to the thermodynamic temperature \(T_{BH}\). The Page curve also shows that the entanglement temperature must be positive after the Page time.

If the Hawking-Page phase transition occurs after the Page time (i.e., \(M_0 > M_c\)), then under the instantaneous transition approximation at \(u = u_{HP}\), the entanglement temperature again tends to zero: \(T_{\text{ent}} \to 0\) at the phase transition point.

\subsection*{Entanglement temperature during evaporation}

To summarize the behavior of the entanglement temperature at different stages of black hole evaporation, we present Table~\ref{tab:ent}. The sign of \(T_{\text{ent}}\) is directly determined by the relative signs of \(\delta E\) and \(\delta S_{\text{gen}}(R)\) through the first law \(\delta E = T_{\text{ent}} \delta S_{\text{gen}}(R)\).

\begin{table}[htbp]
\centering
\caption{Summary of the entanglement temperature \(T_{\text{ent}}\) during black hole evaporation. The sign of \(T_{\text{ent}}\) is determined by the relative signs of \(\delta E\) and \(\delta S_{\text{gen}}(R)\).}
\label{tab:ent}
\begin{tabular}{c|c|c}
\hline
Evolution stage & \(\delta E\) \& \(\delta S_{\text{gen}}\) & \(T_{\text{ent}}\) \\
\hline
Before Page time & \(\delta E<0,\ \delta S>0\) & \(<0\) (negative) \\
At Page time & \(\delta S\to 0\) & diverges \\
After Page time & \(\delta E<0,\ \delta S<0\) & \(>0\) (positive, \(\approx T_{BH}\)) \\
At Hawking-Page transition & both jump & \(\to 0\) \\
\hline
\end{tabular}
\end{table}

As shown in the table, before the Page time the negative sign of \(T_{\text{ent}}\) reflects the information deposition phase. At the Page time, the entropy change passes through zero, causing a divergence in \(T_{\text{ent}}\) (the system passes through an infinite effective temperature). After the Page time, \(T_{\text{ent}}\) becomes positive, signaling the information release phase where the entanglement entropy decreases. Finally, at the Hawking-Page phase transition point (when it occurs), the instantaneous jump in both \(\delta E\) and \(\delta S_{\text{gen}}(R)\) drives the entanglement temperature to zero.

For a reduced density matrix \(\rho_A\) of a subsystem \(A\), one can define a modular Hamiltonian \(K_A\) such that
\begin{equation}
\label{eq:rh0-1}
\rho_A = \frac{e^{-K_A}}{Z}, \qquad Z = \operatorname{Tr}(e^{-K_A}).
\end{equation}
Its eigenvalues constitute the entanglement spectrum, which determines the quantum state structure of subsystem \(A\).

Now define the entanglement temperature \(T_{\text{ent}}\) via
\begin{equation}
\label{eq:rho-Tent}
T_{\text{ent}} = \frac{1}{\beta_{\text{ent}}}, \qquad
\rho_A = \frac{e^{-\beta_{\text{ent}} H_{\text{mod}}}}{Z},
\end{equation}
where the modular Hamiltonian \(K_A\) is identified with \(\beta_{\text{ent}} H_{\text{mod}}\). Here \(H_{\text{mod}}\) is an operator that may differ from the physical Hamiltonian, but whose eigenvalues determine the distribution of the entanglement spectrum.

In terms of the eigenvalues \(\varepsilon_n\) of \(H_{\text{mod}}\), the occupation probability is
\begin{equation}
\label{eq:p-rh0}
p_n = \frac{e^{-\beta_{\text{ent}} \varepsilon_n}}{Z}.
\end{equation}
Hence, when \(T_{\text{ent}} < 0\), the system exhibits population inversion: high entanglement energy levels are more likely to be occupied than low ones. This characterizes the information deposition phase, in which the black hole and the radiation are establishing quantum correlations, and the entanglement entropy increases.

When \(T_{\text{ent}} > 0\), low entanglement energy levels are more occupied than high ones. In quantum statistical mechanics, a negative temperature system is ``hotter'' than any positive temperature system, meaning that energy flows from the negative temperature system to any positive temperature heat bath. This provides a natural physical interpretation for the transition from the information deposition phase (\(T_{\text{ent}} < 0\)) to the information release phase (\(T_{\text{ent}} > 0\)) in black hole evaporation. In the deposition phase, the system is in an ``entanglement excited state'' and the entanglement entropy rises; in the release phase, the system is in an ``entanglement ground state'' and the entanglement entropy decreases.

If the Hawking-Page phase transition occurs before the Page time, then by the time the evolution reaches the phase transition point \(u = u_{\text{HP}}\), all information stored in the black hole is released at once, completing the information escape process. If the Hawking-Page phase transition occurs after the Page time, the system first crosses the Page time, passing through a state of infinite entanglement temperature. After this stage, the system enters the information release phase, releasing energy and information into the heat bath; the entanglement temperature becomes positive. Finally, at the Hawking-Page phase transition point \(u = u_{\text{HP}}\), all remaining information is completely released.

When the Hawking-Page phase transition occurs, a stable black hole phase becomes a thermal radiation phase in the bulk. The latent heat is
\begin{equation}
\delta E^{c} = T_{HP} \, \delta S_{BH}^{(c)} = T_{\text{ent}} \, \delta S_{\text{gen}}(R) = \frac{1}{4G_N}.
\end{equation}
There is a jump in the thermodynamic entropy of the black hole:
\begin{equation}
\delta S_{BH}^{(c)} = \frac{\pi l}{2 G_N},
\end{equation}
and it also corresponds to a jump in the entanglement entropy:
\begin{equation}
\delta S_{\text{gen}}(R) \approx \frac{2c\pi^3 l^3}{3\sigma G_N} \left(1 - \frac{1}{1 + \frac{\sigma T_0 G_N}{\pi^2 l^2}u}\right),
\end{equation}
before the Page time, while after the Page time \(\delta S_{\text{gen}}(R) \approx \delta S_{BH}^{(c)}\). From the above results, we see that when the Hawking-Page phase transition occurs, there is an energy change — the latent heat \(\delta E^{c}\) — which connects the quantum information of the black hole.

These results imply that black hole information does not have to escape solely via Hawking radiation; the Hawking-Page phase transition of the black hole can also cause information to escape.

According to the model of Ref.~\cite{Sun:2018muq}, the thermodynamic phase transition of black holes can be interpreted as a transition between quantized black hole levels. If we interpret the Hawking-Page phase transition as a jump between two levels, the black hole entropy jump from the excited level \(n\) to the ground level \(0\) is
\begin{equation}
S_{BH, n \to 0}^{(c)} = S_{BH, n}^{(c)} - S_{BH, 0}^{(c)}.
\end{equation}
Then one may think that black hole information can also escape via such a transition.

From the above results, we infer that if a black hole undergoes a first-order phase transition, the effect of such a phase transition on the Page curve is universal. Namely, it also applies to higher dimensional and different types of black holes. In the same way, we conclude that black hole information can escape via the transition when a first-order phase transition takes place.

\section{Conclusion}

We have employed the Vaidya metric to construct spacetimes that contain evaporating BTZ black holes. This provides a simple toy model of the entire process: from the gravitational collapse of a null shell that forms a black hole of mass \(M_0\), through its complete evaporation, ending with a pure AdS spacetime.

We compute the entanglement entropy for the spherically symmetric Vaidya model of evaporating BTZ black holes, considering both scenarios with and without island configurations. These results are then used to analyze the Page curve in the context of the black hole information paradox, and to extract the corresponding Page time and scrambling time for our model. In particular, we find that the scrambling time varies with time for evaporating BTZ black holes, which differs from the case of eternal black holes.

We also study the effect of the Hawking-Page phase transition on the Page curve for the evaporating BTZ black hole, both with and without island configurations. The results show that, under certain conditions (e.g., the initial mass \(M(0)=M_0\)), the phase transition can occur before, at, or after the Page time. In particular, if the transition occurs before or at the Page time, it does not require the appearance of an island and yet still reproduces the Page curve behavior.

We calculate the holographic entanglement entropy for the BTZ black hole in the gauge/gravity setup when the Hawking-Page phase transition takes place. We also investigate the first law of entanglement entropy both on the boundary and in the bulk. Since \(\delta M\) is always negative while \(\delta S_{\text{gen}}(R)\) first becomes positive and then negative, the entanglement temperature changes from negative to positive, diverging and flipping sign at the Page time. This reveals that the information deposition phase of black hole evaporation corresponds to a negative entanglement temperature (where the entanglement entropy increases), while the information release phase corresponds to a positive entanglement temperature (where the entanglement entropy decreases). It also shows that the system in the entanglement-accumulating state is ``hotter'' than any positive-temperature system.

If the Hawking-Page phase transition occurs before the Page time, then by the time the evolution reaches the phase transition point, all information stored in the black hole is released at once, completing the information escape process. This indicates that the entanglement entropy accomplishes information release via a quantum jump. If the phase transition occurs after the Page time, the system first crosses the Page time, passing through a state of infinite entanglement temperature. After this stage, the system enters the information release phase, releasing energy and information into the heat bath, and the entanglement temperature becomes positive. Finally, at the Hawking-Page phase transition point, all remaining information is completely released, again implying that the entanglement entropy completes information release via a quantum jump.

Furthermore, we speculate that the effect of first-order phase transitions on black hole information is universal. These results imply that black hole information does not have to escape solely via Hawking radiation; the first-order phase transition itself can also cause the information to be released.

\section*{Acknowledgments}

We would like to thank the National Natural Science Foundation of China~(Grant No.41030001) for supporting us on this work.

 \bibliographystyle{unsrt}
 \bibliography{reference}

\begin{thebibliography}{10}

\bibitem{Hawking1975}
S.~W. Hawking.
\newblock Particle creation by black holes.
\newblock {\em Communications in Mathematical Physics}, 43(3):199--220, Aug
  1975.

\bibitem{Maldacena:1997re}
Juan~Martin Maldacena.
\newblock {The Large N limit of superconformal field theories and
  supergravity}.
\newblock {\em Int. J. Theor. Phys.}, 38:1113--1133, 1999.
\newblock [Adv. Theor. Math. Phys.2,231(1998)].

\bibitem{PhysRevLett.96.181602}
Shinsei Ryu and Tadashi Takayanagi.
\newblock Holographic derivation of entanglement entropy from the anti--de
  sitter space/conformal field theory correspondence.
\newblock {\em Phys. Rev. Lett.}, 96:181602, May 2006.

\bibitem{Page:1993wv}
Don~N. Page.
\newblock {Information in black hole radiation}.
\newblock {\em Phys. Rev. Lett.}, 71:3743--3746, 1993.

\bibitem{Page:2013dx}
Don~N. Page.
\newblock {Time Dependence of Hawking Radiation Entropy}.
\newblock {\em JCAP}, 09:028, 2013.

\bibitem{Almheiri:2019hni}
Ahmed Almheiri, Raghu Mahajan, Juan Maldacena, and Ying Zhao.
\newblock {The Page curve of Hawking radiation from semiclassical geometry}.
\newblock {\em JHEP}, 03:149, 2020.

\bibitem{Penington:2019npb}
Geoffrey Penington.
\newblock {Entanglement Wedge Reconstruction and the Information Paradox}.
\newblock {\em JHEP}, 09:002, 2020.

\bibitem{Almheiri:2019psf}
Ahmed Almheiri, Netta Engelhardt, Donald Marolf, and Henry Maxfield.
\newblock {The entropy of bulk quantum fields and the entanglement wedge of an
  evaporating black hole}.
\newblock {\em JHEP}, 12:063, 2019.

\bibitem{Almheiri:2020cfm}
Ahmed Almheiri, Thomas Hartman, Juan Maldacena, Edgar Shaghoulian, and
  Amirhossein Tajdini.
\newblock {The entropy of Hawking radiation}.
\newblock {\em Rev. Mod. Phys.}, 93(3):035002, 2021.

\bibitem{Almheiri:2019qdq}
Ahmed Almheiri, Thomas Hartman, Juan Maldacena, Edgar Shaghoulian, and
  Amirhossein Tajdini.
\newblock {Replica Wormholes and the Entropy of Hawking Radiation}.
\newblock {\em JHEP}, 05:013, 2020.

\bibitem{Penington:2019kki}
Geoff Penington, Stephen~H. Shenker, Douglas Stanford, and Zhenbin Yang.
\newblock {Replica wormholes and the black hole interior}.
\newblock {\em JHEP}, 03:205, 2022.

\bibitem{PhysRevLett.110.091602}
Jyotirmoy Bhattacharya, Masahiro Nozaki, Tadashi Takayanagi, and Tomonori
  Ugajin.
\newblock Thermodynamical property of entanglement entropy for excited states.
\newblock {\em Phys. Rev. Lett.}, 110:091602, Mar 2013.

\bibitem{Blanco:2013joa}
David~D. Blanco, Horacio Casini, Ling-Yan Hung, and Robert~C. Myers.
\newblock {Relative Entropy and Holography}.
\newblock {\em JHEP}, 08:060, 2013.

\bibitem{Wong:2013gua}
Gabriel Wong, Israel Klich, Leopoldo~A. Pando~Zayas, and Diana Vaman.
\newblock {Entanglement Temperature and Entanglement Entropy of Excited
  States}.
\newblock {\em JHEP}, 12:020, 2013.

\bibitem{PhysRevD.95.106007}
Ki-Seok Kim and Chanyong Park.
\newblock Renormalization group flow of entanglement entropy to thermal
  entropy.
\newblock {\em Phys. Rev. D}, 95:106007, May 2017.

\bibitem{PhysRevD.100.106008}
Ashis Saha, Sunandan Gangopadhyay, and Jyoti~Prasad Saha.
\newblock Holographic entanglement entropy and generalized entanglement
  temperature.
\newblock {\em Phys. Rev. D}, 100:106008, Nov 2019.

\bibitem{Hawking1983}
S.~W. Hawking and Don~N. Page.
\newblock Thermodynamics of black holes in anti-de sitter space.
\newblock {\em Communications in Mathematical Physics}, 87(4):577--588, Dec
  1983.

\bibitem{HENNEAUX1984415}
Marc Henneaux and Claudio Teitelboim.
\newblock The cosmological constant as a canonical variable.
\newblock {\em Physics Letters B}, 143(4):415 -- 420, 1984.

\bibitem{1126-6708-1999-04-024}
Mirjam Cvetic and Steven~S. Gubser.
\newblock Phases of r-charged black holes, spinning branes and strongly coupled
  gauge theories.
\newblock {\em Journal of High Energy Physics}, 1999(04):024, 1999.

\bibitem{Witten:1998zw}
Edward Witten.
\newblock {Anti-de Sitter space, thermal phase transition, and confinement in
  gauge theories}.
\newblock {\em Adv. Theor. Math. Phys.}, 2:505--532, 1998.

\bibitem{TEITELBOIM1985293}
Claudio Teitelboim.
\newblock The cosmological constant as a thermodynamic black hole parameter.
\newblock {\em Physics Letters B}, 158(4):293 -- 297, 1985.

\bibitem{0264-9381-26-19-195011}
David Kastor, Sourya Ray, and Jennie Traschen.
\newblock Enthalpy and the mechanics of ads black holes.
\newblock {\em Classical and Quantum Gravity}, 26(19):195011, 2009.

\bibitem{Arefeva:2021kfx}
Irina Aref'eva and Igor Volovich.
\newblock {A note on islands in Schwarzschild black holes}.
\newblock {\em Teor. Mat. Fiz.}, 214(3):500--516, 2023.

\bibitem{Hashimoto:2020cas}
Koji Hashimoto, Norihiro Iizuka, and Yoshinori Matsuo.
\newblock {Islands in Schwarzschild black holes}.
\newblock {\em JHEP}, 06:085, 2020.

\bibitem{Ling:2020laa}
Yi~Ling, Yuxuan Liu, and Zhuo-Yu Xian.
\newblock {Island in Charged Black Holes}.
\newblock {\em JHEP}, 03:251, 2021.

\bibitem{Chou:2021boq}
Chia-Jui Chou, Hans~B. Lao, and Yi~Yang.
\newblock {Page curve of effective Hawking radiation}.
\newblock {\em Phys. Rev. D}, 106(6):066008, 2022.

\bibitem{Wang:2021woy}
Xuanhua Wang, Ran Li, and Jin Wang.
\newblock {Islands and Page curves of Reissner-Nordstr{\"o}m black holes}.
\newblock {\em JHEP}, 04:103, 2021.

\bibitem{Matsuo:2020ypv}
Yoshinori Matsuo.
\newblock {Islands and stretched horizon}.
\newblock {\em JHEP}, 07:051, 2021.

\bibitem{Saha:2021ohr}
Ashis Saha, Sunandan Gangopadhyay, and Jyoti~Prasad Saha.
\newblock {Mutual information, islands in black holes and the Page curve}.
\newblock {\em Eur. Phys. J. C}, 82(5):476, 2022.

\bibitem{Yu:2021rfg}
Ming-Hui Yu, Cheng-Yuan Lu, Xian-Hui Ge, and Sang-Jin Sin.
\newblock {Island, Page curve, and superradiance of rotating BTZ black holes}.
\newblock {\em Phys. Rev. D}, 105(6):066009, 2022.

\bibitem{Chou:2023adi}
Chia-Jui Chou, Hans~B. Lao, and Yi~Yang.
\newblock {Page curve of AdS-Vaidya model for evaporating black holes}.
\newblock {\em JHEP}, 05:342, 2024.

\bibitem{Guo:2023gfa}
Chang-Zhong Guo, Wen-Cong Gan, and Fu-Wen Shu.
\newblock {Page curves and entanglement islands for the step-function Vaidya
  model of evaporating black holes}.
\newblock {\em JHEP}, 05:042, 2023.

\bibitem{Hubeny:2013dea}
Veronika~E. Hubeny and Henry Maxfield.
\newblock {Holographic probes of collapsing black holes}.
\newblock {\em JHEP}, 03:097, 2014.

\bibitem{Almheiri:2013hfa}
Ahmed Almheiri, Donald Marolf, Joseph Polchinski, Douglas Stanford, and James
  Sully.
\newblock {An Apologia for Firewalls}.
\newblock {\em JHEP}, 09:018, 2013.

\bibitem{VanRaamsdonk:2013sza}
Mark Van~Raamsdonk.
\newblock {Evaporating Firewalls}.
\newblock {\em JHEP}, 11:038, 2014.

\bibitem{Cardoso:2005cd}
Tatiana~R. Cardoso and Antonio~S. de~Castro.
\newblock {The Blackbody radiation in D-dimensional universes}.
\newblock {\em Rev. Bras. Ens. Fis.}, 27:559--563, 2005.

\bibitem{Xu:2019krv}
Hao Xu and Man-Hong Yung.
\newblock {Black hole evaporation in Lovelock gravity with diverse dimensions}.
\newblock {\em Phys. Lett. B}, 794:77--82, 2019.

\bibitem{Ong:2014maa}
Yen~Chin Ong, Brett McInnes, and Pisin Chen.
\newblock {Cold black holes in the Harlow{\textendash}Hayden approach to
  firewalls}.
\newblock {\em Nucl. Phys. B}, 891:627--654, 2015.

\bibitem{Polchinski:2016hrw}
Joseph Polchinski.
\newblock {The black hole information problem.}
\newblock In {\em {Theoretical Advanced Study Institute in Elementary Particle
  Physics}: {New Frontiers in Fields and Strings}}, pages 353--397, 2017.

\bibitem{Hubeny:2009rc}
Veronika~E. Hubeny, Donald Marolf, and Mukund Rangamani.
\newblock {Hawking radiation from AdS black holes}.
\newblock {\em Class. Quant. Grav.}, 27:095018, 2010.

\bibitem{Calabrese:2009ez}
Pasquale Calabrese, John Cardy, and Erik Tonni.
\newblock {Entanglement entropy of two disjoint intervals in conformal field
  theory}.
\newblock {\em J. Stat. Mech.}, 0911:P11001, 2009.

\bibitem{Casini:2009sr}
H.~Casini and M.~Huerta.
\newblock {Entanglement entropy in free quantum field theory}.
\newblock {\em J. Phys. A}, 42:504007, 2009.

\bibitem{Hayden:2007cs}
Patrick Hayden and John Preskill.
\newblock {Black holes as mirrors: Quantum information in random subsystems}.
\newblock {\em JHEP}, 09:120, 2007.

\bibitem{Dolan:2010ha}
Brian~P. Dolan.
\newblock {The cosmological constant and the black hole equation of state}.
\newblock {\em Class. Quant. Grav.}, 28:125020, 2011.

\bibitem{Calabrese_2004}
Pasquale Calabrese and John Cardy.
\newblock Entanglement entropy and quantum field theory.
\newblock {\em Journal of Statistical Mechanics: Theory and Experiment},
  2004(06):P06002, jun 2004.

\bibitem{Hubeny:2013gta}
Veronika~E. Hubeny, Henry Maxfield, Mukund Rangamani, and Erik Tonni.
\newblock {Holographic entanglement plateaux}.
\newblock {\em JHEP}, 08:092, 2013.

\bibitem{Hubeny:2012wa}
Veronika~E. Hubeny and Mukund Rangamani.
\newblock {Causal Holographic Information}.
\newblock {\em JHEP}, 06:114, 2012.

\bibitem{Sun:2018muq}
Dao-Quan Sun, Jian-Bo Deng, Ping Li, and Xian-Ru Hu.
\newblock {Insight into the Microscopic Structure of an AdS Black Hole from the
  Quantization}.
\newblock {\em Class. Quant. Grav.}, 37(1):015008, 2020.

\end{thebibliography}
\end{document}